\documentclass[11pt]{article}  
\usepackage{times,amsmath,amsfonts,amsthm,amssymb,eucal}
\usepackage{array}
\usepackage{color}
\usepackage{enumerate}
\usepackage{epsfig}
\usepackage{mathrsfs}
\usepackage{fancyhdr}
\usepackage{graphicx}
\usepackage{caption}
\usepackage{subcaption}
\usepackage{graphicx}
\usepackage{makeidx}
\usepackage{multirow}

\theoremstyle{plain}
\newtheorem{theorem}{Theorem}
\newtheorem{proposition}{Proposition}
\newtheorem{lemma}{Lemma}
\newtheorem{corollary}{Corollary}
\newtheorem{definition}{Definition}
\newtheorem{assumption}{Assumption}
\newtheorem{remark}{Remark}

\theoremstyle{definition}
\newtheorem{example}{Example}

\newcommand{\E}{\mathrm{E}}
\def\Theorem{\begin{theorem}\sl}
\def\EndTheorem{\end{theorem}}
\def\Proposition{\begin{proposition}\sl}
\def\EndProposition{\end{proposition}}
\def\Lemma{\begin{lemma}\sl}
\def\EndLemma{\end{lemma}}
\def\Remark{\begin{remark}\sl}
\def\EndRemark{\end{remark}}
\def\Corollary{\begin{corollary}\sl}
\def\EndCorollary{\end{corollary}}
\def\Definition{\begin{definition}\sl}
\def\EndDefinition{\end{definition}}
\numberwithin{equation}{section}

\begin{document}
\title{ \textbf{Asymptotic Theory for M-Estimates in Unstable AR($p$) Processes with Infinite Variance Innovations \footnote{Research  supported by a research grant from the \emph{Natural Sciences and Engineering Research Council of Canada (NSERC)}}}}   
\author{ Maryam Sohrabi$^1$  \footnote{E-mail addresses: Maryam.Sohrabi3@carleton.ca (M. Sohrabi)} and  Mahmoud Zarepour$^2$ \footnote{E-mail addresses: mahmoud.zarepour@uottawa.ca (M. Zarepour)}\\
 \small $^1$ School of Mathematics and Statistics, Carleton University,   Ottawa, ON, Canada  \\
\small $^2$Department of Mathematics and Statistics, University of Ottawa,  ON, Canada }
\date{\today}
\maketitle
\pagestyle {myheadings} \markboth {} {Sohrabi and Zarepour: Asymptotic Theory for M-Estimates in Unstable AR($p$) Processes}

\begin{abstract}
In this paper, we present the asymptotic distribution of M-estimators for parameters in unstable AR($p$) processes. The innovations are assumed to be in the domain of attraction of a symmetric stable law with index $0<\alpha\le2$. In particular, in the case of  repeated unit roots or conjugate complex unit roots,  M-estimators have a higher asymptotic rate of convergence compared to the least square estimators and the asymptotic results can be written as It\^{o} stochastic integrals.\\

\vspace{9pt}


\vspace{9pt}

\vspace{9pt}
\end{abstract}
\section {Introduction}
\label{intro}
Consider the autoregressive process of order $p$ (AR($p$))
\begin{align}
\label{1}
 \phi(B)X_t=\epsilon_{t},
\end{align}
where $B$ is the backward operator and
\begin{align}
\label{12}
 \phi(z)=1-\phi_1z-\phi_2z^2-\cdots-\phi_pz^p.
\end{align}
The errors $\{\epsilon_{t}\}$ in \eqref{1} form a sequence of independent and identically distributed (i.i.d.) random variables in the domain of attraction of a symmetric stable law with index $0<\alpha\le2$. The model \eqref{1} is referred to as non-stationary  autoregressive  time series if the characteristic polynomial $\phi(\cdot)$ has at least one root on  the boundary of the unit circle.

It is well known that  the unit root tests are particularly an important tool to classify if a time series is stationary or non-stationary.
When economic variables are non-stationary, estimates  may generate a spurious model unless they are cointegrated. A unit root test can be used for cointegration of two processes. The analysis of unit-root processes and cointegrated time series is likely to be the one of the most important and controversial topics in econometrics in the last few decades.  The case where  innovations are in the domain of attraction of the Gaussian distribution has received considerable attention in  cointegration literature; see for example  Engle and Granger (1987) and  Park and Phillips (1988). However, many empirical studies show that heavy-tailed and asymmetrically distributed samples are frequently observed in economic and, especially, financial time series. In these cases, the  Gaussian models are not applicable. Paulauskas and Rachev (1998)  develop the asymptotic theory for econometric cointegration processes under the assumption of infinite variance innovations with different tail indices.

The problem of conducting asymptotic inference for time series with unit roots has been a challenging topic of interest for some time. In cases where errors (innovations)  have finite variance, Dickey and Fuller (1979), and Phillips and Perron (1988) provide the asymptotic theory for the least squares (LS) estimators in an AR(1) process with one unit root. Chan and Wei (1988) study the large sample theory for a  non-stationary autoregressive AR($p$) model when the innovations form a sequence of martingale differences with respect to an increasing sequence of $\sigma$-fields $\{\mathcal{F}_n\}$.

With infinite variance innovations, Chan and Tran (1989) consider the Dickey-Fuller test when the errors are in the domain of attraction of a stable law. Phillips (1990) extends the results of Chan and Tran (1989) to find the limit theory of the parameters in an AR(1) process with weakly dependent errors in the domain of attraction of a stable law. Since both the Dickey-Fuller  and Phillips-Perron  statistics are based on LS estimation, they do not take advantage of the heavy tails of the innovations and can exhibit rather poor power performance. Thus,  it is important to consider estimation and inference procedures that are robust to departures from finite variance condition. One way to achieve robustness is the use of the M-estimate method. With an appropriate choice of a  loss function, M-estimates have a number of desirable properties when the errors are heavy tailed.
 Knight (1989) considers the asymptotic behavior of the  LS estimates and M-estimates for  the random walk model. The results establish that self-normalized M-estimates are asymptotically normal and their rate of convergence is higher than the LS estimates.  Davis, Knight, and Liu (1992) mention that M-estimates are more appropriate when the distribution of innovations are heavy-tailed. This follows from the fact that M-estimates give less weight to the outliers.  Samarakoon and Knight (2009) develop a class of unit root tests based on M-estimators
in an AR process with a unit root derived by infinite variance innovations.

Chan and Zhang (2012) obtain the limiting distribution for the LS estimates of the parameters for unstable AR($p$) processes, with i.i.d. innovations in the domain of attraction of a stable law. They show that the limiting distribution of the LS estimate is a function of integrated stable processes. However, for model \eqref{1}
with unit roots when $\{\epsilon_{t}\}$  is a sequence of random variables with infinite variance, a complete theory on a more efficient estimating technique is still missing in the literature.

In this paper, we consider an important class of unstable autoregressive time series models with many practical implications. An example of these time series are seasonal models, where $\phi(z)$ may have  several real and complex conjugate roots on the unit circle. We derive the asymptotic distribution of M-estimators for the parameters in an unstable AR($p$) process, where the innovations are in the domain of attraction of a stable law with index $0<\alpha\le2$. Our results show that, similar to the previous cases, M-estimators have higher asymptotic rate of convergence than LS estimators.
This paper is organized as follows. Section 2 provides some necessary preliminary concepts. In Section 3, the limiting distribution of M-estimates in an AR($p$) model is presented and  Section 4 consists of our simulation study. Due to the complexity of the limiting distributions a brief discussion and a bootstrap simulation  scheme are presented in Section 5. We summarize our results in Section 6 and finally, the proof of the main theorem is outlined in Appendix A.

\section{Preliminaries}
Consider the AR($p$) model in \eqref{1} with characteristic polynomial in \eqref{12}. The classical M-estimator $\hat{\Phi}=(\hat{\phi}_{1},\hat{\phi}_{2},\ldots,\hat{\phi}_{p})^\intercal$ of $\Phi=(\phi_1,\phi_2,\ldots,\phi_p)^\intercal$  minimizes
\begin{align*}
\sum_{t=p+1}^{n}\rho(X_t-\beta_{1}X_{t-1}-\cdots-\beta_{p}X_{t-p}),
\end{align*}
with respect to $\left(\beta_1 , \ldots , \beta_p\right)$, where $\rho$ is an almost everywhere differentiable convex function. This guarantees the uniqueness of the solution. For more details see Davis et al. (1992). Usually, $\rho(x)$ grows at a slower rate than $x^2$, as $|x|$ gets large. An example for $\rho(\cdot)$ is the Huber  loss function given by
\begin{align}
\label{Hub}
{\rho _H}\left( x \right)= \tfrac{1}{2}{x^2}I\left(\left| x \right| \le c\right)+\left(c\left| x \right| - {{\tfrac{1}{2}}{c^2}}\right)I\left(\left| x \right| > c\right)
\end{align}
for a known constant $c$, where   $I(\cdot)$ denotes the indicator function. Throughout this paper we impose the  following assumptions on the function $\rho(\cdot)$.
\begin{assumption}
(A1)  Let $\rho$ be a convex and twice differentiable function, and take $\psi=\rho'$.
\end{assumption}
\begin{assumption}
(A2) $\E(\psi(\epsilon_1))=0$ and $\E(\psi^{2}(\epsilon_1))<\infty$.
\end{assumption}
\begin{assumption} (A3)
$0<|\E(\psi'(\epsilon_{1}))|<\infty$  and $\psi'(\cdot)$ satisfies the Lipschitz- continuity condition; i.e., there exists a real constant $k\geqslant 0$ such that for all $x$ and $y$,
 \begin{align*}
|\psi'(x)-\psi'(y)|\leq k|x-y|.
\end{align*}
\end{assumption}
\noindent Note that for the Assumptions A1-A3, sometimes $\rho'$ does not exist everywhere. In this case, although $\rho'$ is not differentiable at a countable number of points, the results will usually hold with some additional complexity in the proofs. Moreover, we assume that the innovations $\{\epsilon_t\}$ satisfy:
\begin{assumption} (A4) The innovations $\{\epsilon_{t}\}$ are i.i.d. random variables in the domain of attraction of a stable law with index $0<\alpha\leq2$. Note that for  $0<\alpha<2$, the innovations have regularly varying tail probabilities as specified by
$$ P(|\epsilon_{1}|>x)=x^{-\alpha}L(x)$$
for some  slowly varying  function $L$ at $\infty$ with $\alpha>0$ as $x\to\infty$,
\begin{align*}
\frac{P(\epsilon_{1}>x)}{P(|\epsilon_{1}|>x)}\to p
\quad,
\frac{P(\epsilon_{1}\le -x)}{P(|\epsilon_{1}|>x)}\to q, \quad, 0\le p\le 1,~ q=1-p.
\end{align*}
\end{assumption}

\vspace{5mm}
\noindent Symmetry is a common assumption for innovations. For $0< \alpha< 1$, symmetry is not required. However, for $1< \alpha\leq2$ we only need $E(\epsilon_1)=0$. For sake of simplicity we assume symmetry on innovations; i.e., $p=q=1/2$; see Assumption A4 implies that:
\begin{align}
\label{4}
S_{n}(t)= a_{n}^{-1}\sum_{k=1}^{[nt]}\epsilon_k\ \overset{d}\to \ S(t) \quad \mbox{in} \quad D[0,1],
\end{align}
where $\overset{d}\to $ denotes convergence in distribution with respect to the Skorohod topology and  $[x]$ stands for integer part of $x$. The Skorohod space of the c\`{a}dl\`{a}g functions defined on $[0,1]$, equipped with the Skorohod topology, is denoted by $D=D[0,1]$.  Here, $\{a_n\}$  is a sequence of positive constants such that
\begin{align}
\label{an}
a_n=\inf\{x:P[|\epsilon_1|>x]\leq n^{-1}\}.
\end{align}
Moreover, $S(\cdot)$ is a stable process and from Resnick and Greenwood (1979) (see also Knight (1989)), the representation for $S$ can be shown to be
\begin{align}
\label{6}
S\left( t \right) =
\left\{
{\begin{array}{*{20}{l}}
{\sum\nolimits_{k = 1}^\infty  {{\delta_k}\Gamma _k^{-1/\alpha }I\left( {{U_k} \leq t} \right)} }&{\rm{if}\,0 < \alpha  < 2,}\\
{\text{standard Brownian motion}}&{\rm{if}\,\alpha  = 2.}
\end{array}} \right.
\end{align}
 Here and throughout this paper, $\{U_{k}\}$ is a sequence of i.i.d. $U[0,1]$ random variables and $\{\delta_k\}$ is a sequence of i.i.d. random variables such that $P(\delta_k=1)=p$, $P(\delta_k=-1)=q$, and $p+q=1$. Also, $\Gamma_1,\Gamma_2,\ldots$ are the arrival times of a Poisson process with Lebesgue mean measure and independent of $\{\delta_{k}\}$. Note that $\{U_{k},\Gamma_{k},\delta_{k}\}$ are mutually independent and the series in \eqref{6} is convergent if either $0<\alpha<1$ or $p=q=1/2$.  For more details, see LePage, Woodroofe, and Zinn (1981).

To derive the main result of this paper, we assume that conditions A1-A4 hold and we  define the following processes on the Skorohod space $D[0,1]$:
{\small
\begin{align}
\label{3}
\nonumber S_{n}^{(1)}(t)   &= a_{n}^{-1}\sum_{k=1}^{[nt]}(-1)^{k}\epsilon_k, \ \ \ \mathbf{T}_{n}(t)= \left({\begin{array}{*{20}{l}}{T_{1,n}(t)} \\
          {T_{2,n}(t)}\end{array}}\right)   = a_{n}^{-1}\sum_{k=1}^{[nt]}\left({\begin{array}{*{20}{l}}{\cos (k\theta)} \\
          {\sin (k\theta)}\end{array}}\right)\epsilon_{k},\\
        \mathbf{R}_{n}(t)  &= \left({\begin{array}{*{20}{l}}{R_{1,n}(t)} \\
          {R_{2,n}(t)}\end{array}}\right)   = {n}^{-1/2}\sum_{k=1}^{[nt]}\left({\begin{array}{*{20}{l}}{\sin\left((k-1)\theta\right)} \\ {\cos\left((k-1)\theta\right)}\end{array}}\right)\psi(\epsilon_{k}),\\
\nonumber  W_{n}(t)   & = n^{-1/2}\sum_{k=1}^{[nt]}\psi(\epsilon_k),\ \ \ V_{n}(t)= n^{-1/2}\sum_{k=1}^{[nt]}\big{[}\psi'(\epsilon_k)-\E\left(\psi'(\epsilon_k)\right)\big{]}.
\end{align}}
 Similar to Theorem 4 of Resnick and Greenwood (1979), we can show that
\begin{align}
\label{5}
\big{(}S_{n}(\cdot),S^{(1)}_{n}(\cdot),\mathbf{T}_{n}(\cdot),W_{n}(\cdot),\mathbf{R}_{n}(\cdot),V_{n}(\cdot)\big{)}^\intercal\
 \overset{d}\to \
\big{(}S(\cdot),S^{(1)}(\cdot),\mathbf{T}(\cdot),W(\cdot), \mathbf{R}(\cdot),V(\cdot)\big{)}^\intercal,
\end{align}
where $S(\cdot)$ and $S^{(1)}(\cdot)$ are stable processes and $\mathbf{T}(\cdot)$ is a bivariate stable process which is defined in Lemma  \ref{Asy.Tn2} (see \eqref{15}). Also, $W(\cdot)$ and $V(\cdot)$ are standard Brownian-motion processes with $\E(W^2(t)) = t\E\left(\psi^2(\epsilon_1)\right)$ and $\E(V^2(t))= t{\rm Var}\left(\psi'(\epsilon_1)\right)$. Finally,  since both
$\small{\sum_{k=1}^{n}\sin^{2}((k-1)\theta)}$ and $\small{\sum_{k=1}^{n}\cos^{2}\left((k-1)\theta\right)}$ are $O_p(n)$ for large values of $n$, the Lindeberg Feller central limit theorem and the tightness of the partial sum process $\mathbf{R}_{n}(\cdot)$ give
\begin{align}
\label{DR}
\mathbf{R}(t)=
\left(
\begin{array}{cc}
 R_1(t) \\
  R_2(t)
\end{array}
\right)=
\frac{\E\left(\psi'^2(\epsilon_1)\right)}{2}
\left(
\begin{array}{cc}
 {\cal W}_1(t) \\
  {\cal W}_2(t)
\end{array}
\right).
\end{align}
Here, ${\cal W}_1(\cdot)$, and ${\cal W}_2(\cdot)$ are independent standard Brownian motion processes.
 For $0 < \alpha < 2$, $\left(S(\cdot), S^{(1)}(\cdot), \mathbf{T}(\cdot)\right)^\intercal$  is independent of $\left(W(\cdot), \mathbf{R}(\cdot),V(\cdot)\right)^\intercal$. The dependence structure can be constructed by applying the continuous mapping theorem on \eqref{Rez1} in Appendix 1. For more details see Resnick and Greenwood (1979).

\section{The limiting distribution for AR($p$)}
The limiting distribution for the M-estimates of the parameters for an infinite-variance random-walk processes is obtained in Knight (1989). To generalize, we extend the results of Knight (1989) to the AR($p$) process when characteristic roots may have different multiplicities and lie on the unit circle. To  derive the asymptotic behavior of the M-estimates, consider the AR($p$) model  in (\ref{1}) when the errors satisfy Assumption A4. Define the process
\begin{align}
\label{Zn}
A_n(q_1,\ldots,q_p)=\sum_{t=p+1}^{n}\left[\rho(\epsilon_t-q_1\mathfrak{b}_{n_1}^{-1}X_{t-1}-\cdots-q_p\mathfrak{b}_{n_p}^{-1}X_{t-p})
-\rho(\epsilon_t)\right],
\end{align}
where $\boldsymbol{q}_n=\mathcal{B}_n(\hat{ \Phi }-\Phi)$ and $\mathcal{B}_n=\text{diag}(\mathfrak{b}_{n_1},\ldots,\mathfrak{b}_{n_p})$ is the matrix of appropriate normalizing constants (see Davis et al. (1992)).  Note that the diagonal entries of $\mathcal{B}_n$ vary according to different roots with different multiplicities and they will be specified in Theorem \ref{Theorem1}. Thus, it is reasonable to expect that the minimizer of the process $A_n$ can be written as
\begin{align}
\label{arp}
\small
 (\mathcal{B}_n^\intercal)^{-1}\left({\hat{ \phi }_{1}}-\phi_1,{\hat{ \phi }_{2}}-\phi_2,\ldots,{\hat{ \phi }_{p}}-\phi_{p}\right)^\intercal=\mathcal{D}_n^{-1}\mathcal{P}_n.
\end{align}
Here
{\small
$$ \mathcal{P}_n=\mathcal{B}_n\left(\sum_{t=p+1}^{n} X_{t-1}\psi(\epsilon_t),\sum_{t=p+1}^{n} X_{t-2}\psi(\epsilon_t),\ldots,\sum_{t=p+1}^{n} X_{t-p}\psi(\epsilon_t)\right)^\intercal$$}
and $\mathcal{D}_n=(d_{i,j}^{(n)})$ is a $p\times p$ matrix such that
\begin{align}
\label{Aarp}
\scriptsize{
\mathcal{D}_n=\mathcal{B}_n
\left(
 \begin{array}{l l l l}
    \sum_{t=p+1}^{n} X_{t-1}^2\psi'(c_t^{(n)}) &\sum_{t=p+1}^{n} X_{t-1} X_{t-2}\psi'(c_t^{(n)}) & \cdots & \sum_{t=p+1}^{n} X_{t-1} X_{t-p}\psi'(c_t^{(n)})\\
   \sum_{t=p+1}^{n} X_{t-1} X_{t-2}\psi'(c_t^{(n)}) & \sum_{t=p+1}^{n} X_{t-2}^2\psi'(c_t^{(n)}) & \cdots &\sum_{t=p+1}^{n} X_{t-2} X_{t-p}\psi'(c_t^{(n)})\\
    \vdots & \vdots & \ddots & \vdots\\
    \sum_{t=p+1}^{n} X_{t-1} X_{t-p}\psi'(c_t^{(n)})  & \sum_{t=p+1}^{n} X_{t-2} X_{t-p}\psi'(c_t^{(n)})  & \cdots &\sum_{t=p+1}^{n} X_{t-p}^2\psi'(c_t^{(n)})
   \end{array}
\right)\mathcal{B}_n^\intercal,}
\end{align}
where  $|c_t^{(n)}-\epsilon_t|\leq|q_1\mathfrak{b}_{n_1}^{-1}X_{t-1}-\cdots-q_p\mathfrak{b}_{n_p}^{-1}X_{t-p}|$.   Asymptotically,  $\psi'(c_t^{(n)})$  can be replaced by $\psi'(\epsilon_t)$ in  \eqref{Aarp}. To see the proof, we assume that $\mathfrak{b}_{n_i}=n^{1/2}a_n$ for all $i=1,\ldots,p$.
Since $|\psi'(\epsilon_{t})-\psi'(c_t^{(n)})|\le k\big{|}\sum_{i=1}^{p}q_{i}n^{-1/2}a_{n}^{-1}X_{t-i}\big{|}$,  we have
\begin{align*}
n^{-1}a_{n}^{-2}\sum_{t=p+1}^{n}X_{t-i}X_{t-j}|\psi'(\epsilon_{t})-\psi'(c_{t}^{(n)})|\le k \sum_{k=1}^p|q_{k}|n^{-1/2}n^{-1}a_{n}^{-3}\sum_{t=p+1}^{n} |X_{t-i}X_{t-j}X_{t-k}|  \overset{P}\to  0.
\end{align*}
In Theorem \ref{Theorem1}, we show that  $n^{1/2}a_n$ is the minimum value for $\mathfrak{b}_{n_i}$, $i=1,\ldots,p$. Thus, this result holds for all other normalizing constants.  For the other entries of $\mathcal{D}_n$, results follow similarly. Furthermore, in \eqref{EPP} (Appendix A)  we  prove that asymptotically each $\psi'(\epsilon_t)$  can be replaced by $\E\left(\psi'(\epsilon_t)\right)$. Therefore, as $n \to \infty$, the matrix $\mathcal{D}_n$ will be nonsingular with probability 1.

The limiting distribution of the parameters in \eqref{arp} is obtained through several steps. First, since time series with different characteristic roots are expected to behave differently,  we can decompose the characteristic polynomial defined in \eqref{12} into the following polynomial:
\begin{align}
\label{32}
\phi(z)=(1-z)^r(1+z)^s\prod_{k=1}^l(1-2\cos(\theta_k)z+z^2)^{d_k}\varphi(z),
\end{align}
where polynomial $\varphi(\cdot)$ corresponds to $q$ roots which are outside the unit circle and $q+r+s+2\sum_{k=1}^ld_k=p$.  Similar to Chan and Wei (1988), we transform $\{X_t\}$ into various components based on the location of their roots. Then we find the limiting behavior of each component individually. Davis et al. (1992) consider the asymptotic behavior of  M-estimate for the parameters in a  stationary AR($p$) process. In this paper we take $\varphi(z)=1$. Define $u_t=\phi(B)(1-B)^{-r}X_t$, $v_t=\phi(B)(1+B)^{-s}X_t$, and $w_t(k)=\phi(B)(1-2\cos(\theta_k)B+B^2)^{-d_k}X_t$ for
$k=1,2,\ldots,l$. Equivalently,
\begin{align*}
\epsilon_t=(1-B)^ru_t=(1+B)^sv_t=(1-2\cos(\theta_k)B+B^2)^{d_k}w_t(k).
\end{align*}
 From Chan and Wei (1988), there exists a nonsingular $p\times p$ matrix $\mathcal{Q}$ such that
\begin{align*}
\mathcal{Q}\mathbf{X_t}=\left(\mathbf{u}_t^\intercal,\mathbf{v}_t^\intercal,\mathbf{w}_t^\intercal(1),\ldots,\mathbf{w}_t^\intercal(l)\right)^\intercal,
\end{align*}
where $\mathbf{X_t}=(X_t,\ldots,X_{t-p+1})^\intercal$, $\mathbf{u_t}=(u_t,\ldots,u_{t-r+1})^\intercal$, $\mathbf{v_t}=(v_t,\ldots,v_{t-s+1})^\intercal$,
and $\mathbf{w_t}(k)=(w_t(k),\ldots,w_{t-2d_k+1}(k))^\intercal$ for  $k=1,2,\ldots,l$. Moreover, let $G_n=\text{diag} \left(J_n,K_n,L_n(1),\ldots,L_n(l)\right)$ be a normalization matrix, where $J_n$ and $K_n$ are as specified  in \eqref{JN} and \eqref{KN} of Appendix A. Also, $L_n(i)$ for $1\leq i \leq l\leq p$ are defined similar to $L_n$ in \eqref{LN} in Appendix A. Then we have
\begin{align*}
G_n\mathcal{Q} \mathbf{X_t} \ \overset{p}\sim \ \text{diag} \left(J_n\mathbf{u_t}, K_n\mathbf{v_t},L_n(1)\mathbf{w_t}(1),\ldots, L_n(l)\mathbf{w_t}(l)\right)
\end{align*}
and
\begin{align}
\scriptsize{
\label{NRF}
\left(\mathcal{Q}^\intercal G_n^\intercal\right)^{-1}(\hat{\Phi}-\Phi)\ \overset{p}\sim \ \left(
 \begin{array}{c}
    (J_n^\intercal)^{-1}\left(\sum_{t=r+1}^{n} \mathbf{u_{t-1}}\mathbf{u_{t-1}}^\intercal\psi'(\epsilon_t)\right)^{-1}\sum_{t=r+1}^{n} \mathbf{u_{t-1}}\psi(\epsilon_t)\\
    (K_n^\intercal)^{-1}\left(\sum_{t=s+1}^{n} \mathbf{v_{t-1}}\mathbf{v_{t-1}}^\intercal\psi'(\epsilon_t)\right)^{-1}\sum_{t=s+1}^{n}
     \mathbf{v_{t-1}}\psi(\epsilon_t)\\
        (L_n(1)^\intercal)^{-1}\left(\sum_{t=2d_1+1}^{n} \mathbf{w_{t-1}}(1)\mathbf{w_{t-1}}^\intercal(1)\psi'(\epsilon_t)\right)^{-1}\sum_{t=2d_1+1}^{n} \mathbf{w_{t-1}}(1)\psi(\epsilon_t)\\
        \vdots\\
    (L_n(l)^\intercal)^{-1}\left(\sum_{t=2d_l+1}^{n} \mathbf{w_{t-1}}(l)\mathbf{w_{t-1}}^\intercal(l)\psi'(\epsilon_t)\right)^{-1}\sum_{t=2d_l+1}^{n} \mathbf{w_{t-1}}(l)\psi(\epsilon_t)
       \end{array}
\right),}
\end{align}
where $a_n \overset{p}\sim b_n$ means $a_n=b_n+o_p(1)$. From \eqref{NRF}, we can find the weak limit behavior of M-estimates for the AR($p$) processes defined in \eqref{1}. The result is presented in  Theorem 1.
\Theorem
\label{Theorem1}
Suppose $\{X_{t}\}$ satisfies \eqref{1} and conditions A1-A4 hold. Then
\begin{align}
\left(\mathcal{Q}^\intercal G_n^\intercal)\right)^{-1}(\hat{\Phi}-\Phi)\ \overset{d}{\to} \ \left((\Gamma^{-1}\mathcal{F})^\intercal,(\Upsilon^{-1}\mathcal{H})^\intercal,(\Lambda_1^{-1}\mathcal{G}_1)^\intercal,\ldots,(\Lambda_l^{-1}\mathcal{G}_l)^\intercal \right)^\intercal,
\end{align}
where $(\Gamma^{-1}\mathcal{F})$ and $(\Upsilon^{-1}\mathcal{H})$, respectively, are defined in \eqref{Lim1} and \eqref{Lim2} in Appendix A. Also,  $(\Lambda_i^{-1}\mathcal{G}_i)$ for  $i=1,\ldots,l$ are given similar to $\Lambda^{-1}\mathcal{G}$ in \eqref{Lim3}. Note that, these limiting distributions are  functional of multiple stochastic integrals of stable processes.
\EndTheorem
\noindent {\textbf{Proof.}} See Appendix A.\\

An illustration of the validity of the results in Theorem 1 is given by  the following example for an AR(2) process.\\
\begin{example}
Suppose $\{X_{t}\}$ is an AR(2) process. Consider the following cases.
\begin{enumerate}[(i)]
  \item When $\phi(z)=1-2z\cos\theta+z^2$,
\begin{align*}
\small
{n^{1/2}}{a_n}
\left(
{\begin{array}{*{20}{c}}
{{\hat{ \phi }_{1}} - 2\cos \theta }\\
{{\hat{ \phi }_{2}} + 1}
\end{array}}
\right)
\ \overset{d}{\to}\
\Gamma _1^{ - 1}
\left(
{\begin{array}{*{20}{c}}
{\frac{{2\sin\theta{\E^{1/2}}\left( {\psi ^2}({\varepsilon _1})\right)\digamma_1 }}{{\E\left( \psi '({\varepsilon _1})\right)\left(\int_{0}^{1}T_1^2(t)d(t)+ \int_{0}^{1}T_2^2(t)d(t)\right) }}}\\
{\frac{{2\sin\theta{\E^{1/2}}\left( {\psi ^2}({\varepsilon _1})\right)\digamma_2 }}{{\E\left( \psi '({\varepsilon _1})\right) \left(\int_{0}^{1}T_1^2(t)d(t)+ \int_{0}^{1}T_2^2(t)d(t)\right)}}}
\end{array}}
\right),
\end{align*}
where
{\small
$$\digamma_1=\cos\theta\left(\int_{0}^{1}T_1(t)dR_1(t)- \int_{0}^{1}T_2(t)dR_2(t)\right)+\sin\theta\left(\int_{0}^{1}T_1(t)dR_2(t)+ \int_{0}^{1}T_2(t)dR_1(t)\right)$$}
and
{\small
$$\digamma_2=\int_0^1T_1(t)dR_1(t)-\int_0^1T_2(t)dR_2(t),\ \ {\Gamma _1} =
\left[
{\begin{array}{*{20}{c}}
1&{\cos \theta }\\
{\cos \theta }&1
\end{array}}
\right].$$}
\item When $\phi(z)=1-2z+z^2$,
\begin{align*}
\small
\left(
{\begin{array}{*{20}{c}}
{{n^{1/2}}{a_n}({\hat{ \phi }_{1}} -2)}\\
{{n^{3/2}}{a_n}({\hat{ \phi }_{1}} - 2)}+{n^{3/2}}{a_n}({\hat{ \phi }_{2}}+1)
\end{array}}
\right)
\ \overset{d}{\to}\
\Gamma _2^{ - 1}
\left(
{\begin{array}{*{20}{c}}
{\frac{{{\E^{1/2}}\left( {\psi ^2}({\epsilon _1})\right) \int_0^1 {S\left( t \right)\,dW\left( t \right)} }}{{\E\left( \psi '({\epsilon _1})\right) }}}\\
{\frac{{\E^{1/2}\left( {\psi ^2}({\epsilon _1})\right) \int_0^1 {\int_0^t {S\left( s \right)\,ds} \,dW\left( t \right)} }}{{\E\left( \psi '({\epsilon _1})\right) }}}
\end{array}}
\right),
\end{align*}
where
\begin{align*}
\small
{\Gamma _2} =
\left(
{\begin{array}{*{20}{c}}
{\int_0^1 {{S^2}\left( t \right)\,dt}}&{\int_0^1 S\left(t\right){\int_0^t {S\left( s \right)\,ds} }\,dt }\\
{\int_0^1 {S\left( t \right)\int_0^t {S\left( s \right)\,ds} \,dt} }&{ \int_0^1 {{{\left( {\int_0^t {S\left( s \right)\,ds} } \right)}^2}\,dt}  }
\end{array}}
\right).
\end{align*}

 \item When $\phi(z)=1+2z+z^2$,
\begin{align*}
\small
\left(
{\begin{array}{*{20}{c}}
{{n^{3/2}}{a_n}({\hat{ \phi }_{1}} + 2)-{n^{3/2}}{a_n}({\hat{ \phi }_{2}} + 1)}\\
{{n^{1/2}}{a_n}({\hat{ \phi }_{2}} + 1)}
\end{array}}
\right)
\ \overset{d}{\to}\
 -\Gamma _3^{ - 1}
\left(
{\begin{array}{*{20}{c}}
{\frac{{{\E^{1/2}}\left( {\psi ^2}({\varepsilon _1})\right) \int_0^1 {\int_0^t {S^{(1)}\left( s \right)\,ds\,} dW\left( t \right)} }}{{\E\left( \psi'({\varepsilon _1})\right) }}}\\
{\frac{{{\E^{1/2}}\left( {\psi ^2}({\varepsilon _1})\right) \int_0^1 {S^{(1)}\left( t \right)\,dW\left( t \right)} }}{{\E\left( \psi'({\varepsilon _1})\right) }}}
\end{array}}
\right),
\end{align*}
where
\begin{align*}
\small
{\Gamma _3} =
\left(
{\begin{array}{*{20}{c}}
{\int_0^1 {{{\left( {\int_0^t {S^{(1)}\left( s \right)ds\,} } \right)}^2}dt} }&{\int_0^1 {S^{(1)}\left( t \right)\int_0^t {S^{(1)}\left( s \right)ds\,dt} } }\\
{\int_0^1 {S^{(1)}\left( t \right)\int_0^t {S^{(1)}\left( s \right)\,ds} \,dt} }&{\int_0^1 {{\left(S^{(1)}\left( t \right)\right)^2}\,dt} }
\end{array}}
\right).
\end{align*}
  \item When $\phi(z)=1-z^2$,
\begin{align*}
\small
\left(
{\begin{array}{*{20}{c}}
{{{n}^{1/2}}{a_{n}}{\hat{ \phi }_{1}}/2 + {{n}^{1/2}}{a_{n}}({\hat{ \phi }_{2}} - 1)}/2\\
{{{n}^{1/2}}{a_{n}}{\hat{ \phi }_{1}}/2 - {{n}^{1/2}}{a_{n}}({\hat{ \phi }_{2}} - 1)}/2
\end{array}}
\right)
\ \overset{d}{ \to}\
\left(
{\begin{array}{*{20}{c}}
{\frac{{{\E^{1/2}}\left( {\psi ^2}({\varepsilon _1})\right) \int_0^1 {S(t)\,dW\left( t \right)} }}{{\E\left( \psi '({\varepsilon _1})\right) \int_0^1 {{S^2}(t)\,dt} }}}\\
{-\frac{{{\E^{1/2}}\left( {\psi ^2}({\varepsilon _1})\right) \int_0^1 {S^{(1)}(t)\,dW\left( t \right)} }}{{\E\left( \psi '({\varepsilon _1})\right) \int_0^1 {{\left(S^{(1)}\left( t \right)\right)^2}\,dt} }}}
\end{array}}
\right),
\end{align*}
\end{enumerate}
where  $S(\cdot)$, $S^{(1)}(\cdot)$,  $\mathbf{T}(\cdot)$, $W(\cdot)$, and $\mathbf{R}(\cdot)$ are  defined  in \eqref{5}.\\
\end{example}

\begin{remark}
Example 1  shows a higher convergence rate of  the M-estimates  compared to the  classical  LS estimates in Chan and Zhang (2012). For instance, the rate of convergence of the M-estimates in case (i) is ${n^{1/2}}{a_n}$ which is significantly higher than that of  LS estimate at the rate of $n$, specially for small values of $\alpha$. Moreover, the results in Example 1, cases (ii) and (iii), show  different  convergence rates for the estimators. Broadly speaking, the linear combination of $\hat{\phi}_{1}$ and $\hat{\phi}_{2}$ converges faster than each of them individually; see
Chan and Zhang (2012).
\end{remark}

\section{Simulation}
\begin{figure}
  \centering
  \includegraphics[width=0.5\textwidth]{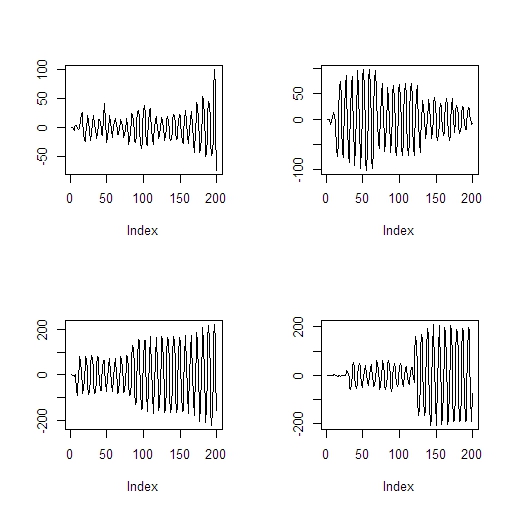}
 \caption{Different sample paths for model \eqref{123} when $\alpha=1.3$ and $n=500$.}
         \label{fig:Complex}
\end{figure}

To investigate the results given in Theorem 1, we carry out a simulation for  model \eqref{1}, when $p=2$ and $\phi(B)=1-\phi_1B-\phi_2B^2$ has complex conjugate unit roots. In other words, we illustrate the asymptotic result of  case (i) in  Example 1  by simulating the following AR(2) process
\begin{align}
\label{123}
X_t=2\cos\theta X_{t-1}- X_{t-2}+\epsilon_t,
\end{align}
where the innovations are i.i.d. symmetric $\alpha$-stable random variables. Figure \ref{fig:Complex} shows different sample paths of Model \eqref{123} when $\alpha=1.3$ and $n=500$.  For the sake of brevity, we only present the M-estimates for $\phi_{1}$ when $\theta=\pi/4$. The simulations for the other unstable cases are similar.

In our simulation study, we consider  the time series $\{X_t\}_{t=0}^n$ in model \eqref{123}, for $n=10$, $20$, $30$, $40$, and $50$  with $\alpha=0.5$, $1$, $1.3$, $1.7$ and $2$. The  time series  are generated 10,000 times for each choice of $n$ and $\alpha$. The M-estimates of $\phi_1$ in AR(2) are calculated for each replicate. Here,  we use the Huber loss function, $\rho_H(x)$ in \eqref{Hub}, with $c=5$ in all cases. The sample median and the sample  90\% inter-percentile range (IPR)  (95th percentile - 5th percentile) for $|\hat{\phi}_{1}-2\cos\theta|$  are tabulated in Table 1. Furthermore,  the simulated results of the sample median and 90\% IPR  for $|\hat{\phi}_{1}-2\cos\theta|$ with the LS method are shown in Table 2.
As seen in Tables 1 and 2, the M-estimates  are significantly  closer to the actual values as $n$ gets larger compared to the  LS estimates, especially  for small values of $\alpha$. Note that, when $\alpha=2$, the  rate of convergence of both M-estimate and  LS estimate are the same.  Moreover,  the 90\% IPR value for the M-estimates are significantly smaller than those for the LS estimates.  The tables confirm that M-estimates are more precise than LS estimates.
\begin{table}[!h]
\centering
\small
\caption{
Median and 90\% IPR (in parentheses) for $|\hat{\phi}_{1}-2\cos\theta|$ in model \eqref{123} by the M-estimate method using the Huber loss function}
\begin{tabular}{lccccc}
  \hline
  {} & {} & {} & Index of stability $\alpha$ & {} & {} \\
  \hline
  $n$ & 0.5 & 1 & 1.3 & 1.7 & 2 \\
  \hline
  10 & 0.0258(0.4141) & 0.1170(0.4097) & 0.1501(0.4048) & 0.1839(0.4003) & 0.1947(0.3961)\\
  20 & 0.0030(0.0815) & 0.0342(0.2494) & 0.0552(0.2981) & 0.0734(0.3456) & 0.0852(0.3673)\\
  30 & 0.0009(0.0245) & 0.0180(0.1367) & 0.0328(0.1797) & 0.0444(0.2133) & 0.0538(0.2323)\\
  40 & 0.0004(0.0107) & 0.0115(0.0841) & 0.0216(0.1234) & 0.0329(0.1628) & 0.0391(0.1788)\\
  50 & 0.0002(0.0063) & 0.0083(0.0584) & 0.0159(0.0916) & 0.0263(0.1278) & 0.0318(0.1396)\\
  \hline
\end{tabular}
\end{table}

\begin{table}[!h]
\centering
\small
\caption{
Median and 90\% IPR (in parentheses) for $|\hat{\phi}_{1}-2\cos\theta|$ in model \eqref{123} by the LS estimate method}
\begin{tabular}{lccccc}
  \hline
  {} & {} & {} & Index of stability $\alpha$  & {} & {} \\
  \hline
  $n$ & 0.5 & 1 & 1.3 & 1.7 & 2\\
  \hline
  10 & 0.0824(2.0654) & 0.1439(0.9341)& 0.1633(0.7900) & 0.1867(0.7794) & 0.194(0.7595)\\
  20 & 0.0297(0.3620) & 0.0543(0.3295) & 0.0660(0.3382) & 0.0765(0.3541) & 0.0852(0.3674)\\
  30 &  0.0180(0.2038) & 0.0351(0.2224) & 0.0429(0.2200) & 0.0477(0.2223) & 0.0538(0.2323)\\
  40 & 0.0133(0.1323) & 0.0257(0.1567) & 0.0311(0.1622) & 0.0362(0.1719) & 0.0391(0.1788)\\
  50 & 0.0105(0.1070) & 0.0202(0.1247) & 0.0243(0.1259) & 0.0286(0.1351) & 0.0318(0.1397)\\
   \hline
\end{tabular}
\end{table}

\section{Bootstrapping}
The asymptotic distributions for the proposed estimator of $\Phi=(\phi_1,\ldots,\phi_p)^\intercal$  by  Chan and Zhang (2012) and this article are generally intractable. Due to the complexity of the  limiting distributions, to make inferences based on $\hat{\phi}_1,\ldots,\hat{\phi}_p$, one may consider a resampling scheme.
 In this section, we briefly suggest using the result of Theorem 1 of Moreno and Romo (2012) and  Theorem 3.1 of Zarepour and Knight (1999a) in model \eqref{1}, when $\{\epsilon_t\}$ is a sequence of i.i.d. random variables and belongs to the domain of attraction of a stable law with index $\alpha$.  Define the point process $\xi_n=\sum\limits_{k=1}^{n} \varepsilon_{\left({k/n},a_n^{-1}\epsilon_k\right)}$,  where $\varepsilon_x$ is a measure defined by $\varepsilon_x(A)=I(x \in A)$ for any Borel set A such that $A\subseteq [0,1]\times \mathbb{R}$. From Resnick (1987), we have
\begin{align}
\label{Rez1}
\sum\limits_{i=1}^{n} \varepsilon_{\left({i/n},a_n^{-1}\epsilon_i\right)}
\ \overset{d}{ \to}\
\sum\limits_{i=1}^{\infty} \varepsilon_{\left(U_i,\delta_i\Gamma_i^{-1/\alpha}\right)},
\end{align}
where $U_{i}\overset{i.i.d.}\sim U[0,1]$ is independent from $\{\delta_i,\Gamma_i\}$, which are defined in \eqref{6}. Here $\overset{d}{ \to}$ indicates weak convergence with respect to vague topology, see Resnick (1987). Now,  assume that $\{\epsilon_1^*,\epsilon_2^*,\cdots,\epsilon_n^*\}$ is an i.i.d. sample from $F_n(x)=\frac{1}{n}\sum_{i=1}^{n}I(\epsilon_i\leq x)$. Then, the bootstrapped point process
\begin{align}
\label{BO}
\sum_{i=1}^{n}\varepsilon_{\left({i/n},a_n^{-1}\epsilon_i^*\right)}\left([0,t]\times\cdot\right)\ \overset{d}\to \ \sum_{i=1}^{\infty}\mathcal{P}_i^*(t)\varepsilon_{\left(U_i,\delta_i\Gamma_i^{-1/\alpha}\right)}(\cdot)
\end{align}
in distribution. Here, $\{\mathcal{P}_i^*(t)\}$ is a sequence of independent Poisson processes with Lebesgue mean measure and $\{\delta_i,\Gamma_i\}$ are the same as before. Note that in this case, the regular bootstrap  asymptotically fails for the bootstrapped point process since the limit in  \eqref{BO} contains the extra Poisson point process $\mathcal{P}_i^*(t)$.
Subsampling of size $m$ when $m/n\rightarrow 0$  as $n\rightarrow \infty$ resolves the asymptotic failure of the regular bootstrap; see Zarepour and Knight (1999b).
By the same argument used for the stationary AR($p$) processes with infinite variance in Davis and Wu (1997), we can prove  that the results of the limiting distributions are asymptotically valid when we use the bootstrap scheme with $m=o(n)$ resampling sample size. Given $\left(X_1 ,\ldots , X_n\right)$, we find M-estimates of $\Phi$ in model \eqref{1} using the objective function in \eqref{Hub}. The residuals are calculated from
$e_{t}=X_t-\hat{\phi}_1X_{t-1}-\cdots-\hat{\phi}_pX_{t-p}, \text{ }t=p+1,2,\ldots,n$. Then,  the bootstrap replicate $\left(X_1^* ,\ldots , X_m^*\right)$ are generated from $X_t^*=\hat{\phi}_1X_{t-1}^*+\cdots+\hat{\phi}_pX_{t-p}^*+{e}_t^*$,
where ${e}_1^*,\ldots,{e}_{m}^*$ is  a sample of size $m$ from $\hat{F}_n(\cdot)=\frac{1}{n-p}\sum_{i=p+1}^{n}I(e_i-\bar{e}\leq \cdot)$, where $\bar{e}=\frac{1}{n-p}\sum_{i=p+1}^{n}e_{i}$. The bootstrap replicate of $\hat{\Phi}$ is found by minimizing
\begin{align*}
A_{m}^*(q_1^*,\ldots,q_p^*)=\sum_{t=p+1}^{m}\left[\rho(e_t^*-q_1^*\mathfrak{b}_{m_{1}}^{-1}X_{t-1}^*-\cdots-q_p^*\mathfrak{b}_{m_{p}}^{-1}X_{t-p}^*)
-\rho(e_t^*)\right],
\end{align*}
with the minimum given by  $\hat{\boldsymbol{q}}_{m}^*=\mathcal{B}_{m}\left(\hat{ \Phi }_{m}^*-\hat{\Phi}\right)$, where $\mathcal{B}_{m}=\text{diag}\left(\mathfrak{b}_{m_{1}},\ldots,\mathfrak{b}_{m_{p}}\right)$ is the matrix of appropriate normalizing constants. The following lemma helps to derive the limiting distribution of bootstrap estimates.
\begin{lemma}
 \label{SRS}
 Let $\{e_1^*,\ldots,e_m^*\}$ be an i.i.d. sample from $\hat{F}_n$  and $\E^*$ denotes the expectation under $\hat{F}_n$. Also, under conditions A1-A4 and  with the subsampling of size $m$ where $m\rightarrow \infty$ and $m/n\rightarrow 0$, we have
\begin{enumerate}[(i)]
\item $S_m^*(\cdot)=a_{m}^{-1}\sum_{i=1}^{[m\cdot]}e_i^*\ \overset{d}\to\  S(\cdot)$ in probability, where $S(\cdot)$ is the stable process defined in \eqref{6},
\item $\E^*(\psi(e_{1}^*))=0$,
\item $m^{-1/2}\hat{\sigma}^{-1}\sum_{i=1}^{[m\cdot]}\psi(e_{i}^*)\overset{d}\to W(\cdot) \ \text{in probability}$,
where $W(\cdot)$ is a standard Brownian motion independent of $S(\cdot)$ and $\hat{\sigma}^2=\E^*(\psi^2(e_{1}^*))=\frac{1}{n-p}\sum_{i=p+1}^{n}\psi^2(e_{i})\overset{p}\to \E(\psi^2(\epsilon_{1}))$,
\item $\E^*(\psi^{'}(e_{1}^*))\overset{p}\to \E(\psi^{'}(\epsilon_{1}))$.
\end{enumerate}
\end{lemma}
We omit the proof of Lemma \ref{SRS} since it follows from a similar argument used in Proposition 4 of  Moreno and Romo (2012) and  Arcones and Gin\'{e} (1989). See also Sohrabi and Zarepour (2016). From Lemma \ref{SRS}, we can conclude that
\begin{align*}
\big{(}S_{m}^*(\cdot),S^{*(1)}_{m}(\cdot),\mathbf{T}_{m}^*(\cdot),W_{m}^*(\cdot),\mathbf{R}_{m}^*(\cdot),V_{m}^*(\cdot)\big{)}^\intercal\
 \overset{d}\to \
\big{(}S(\cdot),S^{(1)}(\cdot),\mathbf{T}(\cdot),W(\cdot), \mathbf{R}(\cdot),V(\cdot)\big{)}^\intercal
\end{align*}
in probability. Here $S_{m}^*(\cdot)$ is defined  as in Lemma \ref{SRS} and $\left\{S^{*(1)}_{m}(\cdot),\mathbf{T}_{m}^*(\cdot),W_{m}^*(\cdot),\mathbf{R}_{m}^*(\cdot),V_{m}^*(\cdot)\right\}$ is defined in \eqref{3} by replacing $\epsilon_i$ by $e_i^*$ and $n$ by $m$. Also, $\left\{S(\cdot),S^{(1)}(\cdot),\mathbf{T}(\cdot),W(\cdot), \mathbf{R}(\cdot),V(\cdot)\right\}$ is as specified in \eqref{5}. Similar to Theorem 3.1 of Davis and Wu (1997), along with applying the continuous mapping theorem,  and  Lemma \ref{SRS} we have
\begin{align*}
\left(Q^\intercal G_m^\intercal\right)^{-1}(\hat{\Phi}^*-\hat{\Phi})\ \overset{w}{\to}\  \left((\Gamma^{-1}\mathcal{F})^\intercal,(\Upsilon^{-1}\mathcal{H})^\intercal,(\Lambda_1^{-1}\mathcal{G}_1)^\intercal,\ldots,(\Lambda_l^{-1}\mathcal{G}_l)^\intercal \right)^\intercal
\end{align*}
in probability, where $(\Gamma^{-1}\mathcal{F})$, $(\Upsilon^{-1}\mathcal{H})$, and  $(\Lambda_i^{-1}\mathcal{G}_i)$ for  $i=1,\ldots,l$ are specified as in Theorem \ref{Theorem1}. For a detailed proof of the asymptotic validity of the $m$ out of $n$ bootstrap in a special case see Sohrabi and Zarepour (2016). Also see Zarepour and Knight (1999a, b), Moreno and Romo (2012), Davis and Wu (1997), and Sohrabi (2016) for some similar techniques. An illustration of the validity of the bootstrap scheme is given by a simulation study in the following example.\\

\begin{example}
We consider bootstrapping for the  parameter $\phi_{1}$  in model  \eqref{123} when $\theta=\pi/4$.
We generate the time series $\{X_t\}_{t=0}^n$ in model \eqref{123}, for the actual sample sizes $n=50$, $100$,  and $200$  with $\alpha=1.3$, and $1.7$ (the cases with $\alpha>1$ are of practical interest). The behavior of the other unstable cases are more or less similar. Then we  implement the following algorithm for each choice of $n$ and $\alpha$.
\begin{enumerate}[(i)]
\item We find M-estimates of the parameters $\phi_1$ and $\phi_2$ in model \eqref{123} using the Huber loss function with $c=5$. Then take
$$\hat{\epsilon}_t=X_t-\hat{\phi}_1 X_{t-1}-\hat{\phi}_2X_{t-2}.$$
\item We draw a sample of size $m$ from centered residuals denoted by  $\hat{\epsilon}_1^*,\ldots,\hat{\epsilon}_m^*$, and find  $\{X_t^*\}_{t=0}^m$ from \eqref{123}. Then we  estimate the parameter $\phi_1$ using the bootstrap observations  by  the same minimization of the objective function used in  step (i).
\item  We repeat  step (ii) $B=3,000$ times to get $\hat{\phi}_1^{1*},\ldots,\hat{\phi}_1^{3,000*}$.  To find a  naive 95\% confidence interval for $\phi_{1}$, we obtain the 2.5th and 97.5th percentiles  of   3,000 bootstrap estimates  as  the lower and upper bound of our confidence interval.
\end{enumerate}
In order to compute the coverage rate of the bootstrap confidence intervals, the original  time series are generated 10,000 times for each choice of $n$ and $\alpha$. Then by applying (i)-(iii), the naive  95\% bootstrap confidence interval for $\phi_{1}$ is calculated for each replicate.  Moreover, to study  how the selection of the resampling size would affect our estimation, we perform the second step with three different resampling sizes $m=n/\ln(\ln(n))$, $n^{0.9}$,  and $n^{0.95}$.
 Finally, the resulting  coverage percentages of the naive $95\%$  bootstrap confidence intervals for the parameter $\phi_{1}$  for different values of $m$, $n$, and $\alpha$ are presented in Table 3. This table shows that the coverage percentages of the naive 95\% bootstrap confidence intervals for $\phi_{1}$ are very close to 95\%. This illustrates that the bootstrap  scheme with $m=o(n)$ resampling sample size is  approximately valid when we have a non-stationary time series with innovations in the domain of attraction of a stable law. The simulation procedure shows $m=n^{0.95}$ performs consistently well in all our cases.

\begin{table}[!h]
\centering
\caption{
Coverage for the naive $95\%$ bootstrap confidence interval for $\phi_{1}$ in  model \eqref{123} }
\label{TC7}
\begin{tabular}{lcccccccccc}
  \hline
  $\alpha$ & {} &  {} & 1.3 &  {} &  {} & {} &  {} & 1.7 & {}\\ \cline{3-5} \cline{4-5} \cline{8-10}
  $n$ & {} & 50 & 100 & 200 & {} &  {} & 50 & 100 & 200\\
  \hline
  $m=n/\ln(\ln(n))$ & {} & 96.1\% & 96.5\% & 97.4\%  & {} &  {} & 95.1\% & 96.5\% & 96.8\%\\
  $m=n^{(0.9)}$ & {} & 97.2\% & 97.4\% & 97.7\%  & {} &  {} & 96.6\% & 96.9\% & 97.4\%\\
  $m=n^{(0.95)}$ & {} & 95.7\% & 96.6\% & 96.3\%  & {} &  {} & 94.0\% & 94.8\% & 94.8\%\\
  \hline
\end{tabular}
\end{table}
\end{example}

\section{Conclusion}
In this paper, the asymptotic properties of the M-estimate have been thoroughly studied  for the unstable AR($p$) processes when $\{\epsilon_{t}\}$ is a sequence of random variables in the domain of attraction of a stable law with index $0<\alpha\leq2$.  The behavior of  these time series   with several real and complex conjugate roots on the unit circle, is completely different from processes with a single unit root. The robust M-estimate method has been used to drive the limiting distribution of estimates of the parameters in the AR($p$) model defined in \eqref{1}.  Although the M-estimates have a better rate of convergence compared to LS estimates, the limiting distributions are not computationally tractable. To remedy this difficulty, we suggest a valid $m$ out of $n$ bootstrap where $m/n\to 0$ as $n\to \infty$. Our simulation study proves that the resampling bootstrap scheme is valid for our non-stationary time series when errors are in the domain of attraction of a stable law.

Based on our analysis, when errors are heavy tailed, M-estimators are always superior to LS estimators. This superiority holds for LAD but some extra conditions should be imposed on innovations.  In this case, Knight (1989) and Davis et al. (1992) assume that the innovations need to have 0 median and a density with respect to Lebesgue measure. Since the computational complexity of both M-estimates and LAD estimates are similar, we do not consider the LAD estimate to avoid imposing extra conditions.

\section{Appendix A}
 To facilitate the proof of Theorem 1, we first present the following two lemmas. The  proof of  Lemma \ref{Asy.Tn1} is similar to the proof given in Theorem 1 of Banjevic, Ishwaran and Zarepour (2002).
\Lemma \label{Asy.Tn1}
Let $(Z_{1,k},Z_{2,k})^\intercal$ be a sequence of symmetric i.i.d. random vectors on $\mathcal{Z}_2=\{(z_1,z_2)^\intercal:z_1^2+z_2^2=1\}$ with a probability distribution $Q$  on the boundary of the unit circle, and
$${\cal X}=\left(\sum\nolimits_{k=1}^{\infty}{Z_{1,k}}{\Gamma_k^{-1/\alpha}},
 \sum\nolimits_{k=1}^{\infty}{Z_{2,k}}{\Gamma_k^{-1/\alpha}}\right)^\intercal.$$
 Then the characteristic function for ${\cal X}$ is $\phi(\mathbf{s})=\exp\left[-K\E\left(|s_1Z_{1,1}+s_2Z_{2,1}|^\alpha\right)\right]$,
where $\mathbf{s}=(s_1,s_2)$, and $K$ is given by
\begin{align*}
\scriptsize
K =
\left\{
{\begin{array}{*{20}{l}}
{\cos (\pi \alpha /2)\Gamma (1 - \alpha )}&{\rm{if}\,0 < \alpha  < 1,}\\
{\pi (2 - \alpha )/(2\alpha )}&{\rm{if}\,\alpha  = 1,}\\
{\cos (\pi \alpha /2)\Gamma (3 - \alpha )/({\alpha ^2} - \alpha )}&{\rm{if}\,1 \le \alpha  < 2.}
\end{array}}
\right.
\end{align*}
where  $\{\Gamma_k\}$ is  defined in \eqref{6} and  is independent from
 $\{Z_{i,k}\}$, $i=1,2$.
\EndLemma
\begin{lemma}
\label{Asy.Tn2} Let $\mathbf{T}_n$ be defined by  \eqref{3}, then
$$\mathbf{T}_n (\cdot)\overset{d}{ \to} \mathbf{T}(\cdot),$$
where $\mathbf{T}(\cdot)=\left(T_1(\cdot),T_2(\cdot)\right)^\intercal$ is a bivariate stable process with index $\alpha$.
\EndLemma
\noindent {\textbf{Proof.}} Consider the point process convergence in \eqref{Rez1}. By applying the continuous mapping theorem, for any $\theta\in(0,2\pi)$, we have
{\small
\begin{align*}
\sum\limits_{k=1}^{n} \varepsilon_{\left(\cos\left(\theta{k/n} \right),\sin\left(\theta{k/n} \right),a_n^{-1}\epsilon_k\right)}
\ \overset{d}{ \to}\
\sum\limits_{k=1}^{\infty} \varepsilon_{\left(\cos\left(\theta U_k\right),\sin\left(\theta U_k\right),\delta_k\Gamma_k^{-1/\alpha}\right)}.
\end{align*}}
Similar to Resnick's (1986) Proposition 3.4, it can be shown that
{\small
\begin{align}
\label{15}
a_n^{-1}\sum\limits_{k=1}^{[nt]}\left(\cos\left(\theta{k/n}\right),\sin\left(\theta{k/n}\right)\right) \epsilon_k
\ \overset{d}{ \to}\
\sum\limits_{k=1}^{\infty}\left(\cos\left(\theta U_k\right),\sin\left(\theta U_k\right)\right)\delta_k\Gamma_k^{-1/\alpha}I(U_k\leq t).
\end{align}}
Notice that we can easily find the finite dimensional distribution for the limiting process in \eqref{15}. For instance, for $t=1$ and by applying  Lemma  \ref{Asy.Tn1} where $\left(Z_{1,k},Z_{2,k}\right)=\left(\cos(\theta U_k)\delta_k,\sin(\theta U_k)\delta_k\right)$, and $p=q=1/2$, we have
{\small
$$\phi(\mathbf{s})=\exp\left[-K \int_{0}^1|s_1\cos(\theta u)+s_2\sin(\theta u)|^\alpha du\right].$$}
$\hfill\square$

\noindent {\textbf{Proof of Theorem 1 with Roots equal to 1.}} First, we consider the unit root  model
\begin{align*}
(1-B)^ru_{t}(\cdot)=\epsilon_{t}.
\end{align*}
To avoid singular limiting distributions, similar to Chan and Wei (1988), we define
\begin{align}
\label{newdif}
u_{t}(j)=(1-B)^{r-j}u_{t} \ \ \mbox{for} \ j=1,2,\ldots,r,
\end{align}
or equivalently $u_{t}(1)=\sum_{j=1}^{t}\epsilon_{j}$ and $u_{t}(j)=\sum_{k=1}^{t}u_{k}(j-1)$ for $j=2,\ldots,r$. From  definition \eqref{4}, we have
$$a_{n}^{-1}u_{[nt]}(1)=S_{n}(t)=:\mathcal{S}_{1,n}(t) \ \overset{d}\to \ S(t)=:\mathcal{S}_{1}(t).$$
Since $u_{t}(j)=\sum_{k=1}^{t}u_{k}(j-1)$ for $j=2,\ldots,r$, the continuous mapping theorem implies that
\begin{align}
\label{gj}
a_{n}^{-1}n^{-(j-1)}u_{[nt]}(j)=\int_{0}^{t}\mathcal{S}_{j-1,n}(s)ds=:\mathcal{S}_{j,n}(t) \ \overset{d}\to \mathcal{S}_{j}(t)
\end{align}
for $j=2,\ldots,r$. We also define
\begin{align}
\label{JN}
J_n=N_n^{-1}C,
\end{align}
where
\begin{align*}
\scriptsize{
C=
\left(
 \begin{array}{c c c c c}
    1 & 0 & 0 & \cdots & 0\\
    1 & -1& 0 & \cdots & 0\\
    \vdots & \vdots& \vdots & \ddots & \vdots\\
    1 & (-1)\binom{r-1}{1} & (-1)^2\binom{r-1}{2} & \cdots & (-1)^{r-1}\\
     \end{array}
\right)}
\end{align*}
and the appropriate normalizing constant is $N_n=\text{diag}\left(n^{r-1/2}a_{n}, n^{(r-1)-1/2}a_{n},\ldots,n^{1/2}a_{n}\right)$.
First, notice that $C\left(\sum_{t=r+1}^{n} \mathbf{u_{t-1}}\mathbf{u_{t-1}}^\intercal\psi'(\epsilon_t)\right)C^\intercal=\Pi$, where the $(i,j)^{th}$ entry of $\Pi$ is
$$\pi_{i,j}=\sum_{t=r+1}^{n} u_{t-1}\left(r-(i-1)\right) u_{t-1}\left(r-(j-1)\right)\psi'(\epsilon_t) \ \ \text{for} \  i,j=1,2,\ldots,r.$$
The joint behavior of $\Pi$ can be studied through  some tedious calculations. Therefore, for simplicity, we only calculate the limiting behavior of each term of $\Pi$  individually by using the following steps. As discussed before,  each $\psi'(\epsilon_t)$  can be replaced by $\E\left(\psi'(\epsilon_t)\right)$ when $n \to \infty$. To see this, note that
{\small
\begin{align}
\label{CAC}
\nonumber\sum_{t=r+1}^nu_{t-1}(i)u_{t-1}(j)\psi'(\epsilon_t)&=\sum_{t=r+1}^nu_{t-1}(i)u_{t-1}(j)\left[ \psi'(\epsilon_t)- {\E}\left(\psi'(\epsilon_t)\right)+ {\E}\left(\psi'(\epsilon_t)\right)\right]\\
 &=\sum_{t=r+1}^nu_{t-1}(i)u_{t-1}(j)\left[\psi'(\epsilon_t)- {\E}\left(\psi'(\epsilon_t)\right)\right]
 +  {\E}\left(\psi'(\epsilon_1)\right)\sum_{t=r+1}^nu_{t-1}(i)u_{t-1}(j)
\end{align}}
for $i,j=1,2,\ldots,r$.  Notice that by \eqref{3} and \eqref{gj}, the first  term on the right side of \eqref{CAC} is $O_p (a_n^{2}n^{(i+j-3/2)})$. Then we have
\begin{align}
\label{EPP}
a_n^{-2}n^{-(i+j-1)}\sum_{t=r+1}^{n} u_{t-1}(i) u_{t-1}(j)[\psi'(\epsilon_t)-\E\left(\psi'(\epsilon_t)\right)]\ \overset{p}\to \ 0.
\end{align}
From \eqref{gj}, along with applying the continuous mapping theorem  and Proposition 2 of Paulauskas and Rachev (1998), we can find the limiting of each term of the matrix $\Pi$ where
\begin{align*}
a_n^{-2}n^{-(i+j-1)}\sum_{t=r+1}^{n} u_{t-1}(i) u_{t-1}(j)\psi'(\epsilon_t)\ \overset{d}\to \  \E\left(\psi'(\epsilon_1)\right)\int_0^1\mathcal{S}_{i}(t)\mathcal{S}_{j}(t)dt \ \ \mbox{for} \ i,j=1,2,\ldots,r.
\end{align*}
Therefore,
$$ J_n\left(\sum_{t=r+1}^{n} \mathbf{u_{t-1}}\mathbf{u_{t-1}}^\intercal\psi'(\epsilon_t)\right)J_n^{\intercal} \ \overset{d}\to \ \Gamma,$$
where  $\Gamma=(\gamma_{i,j})$ is  a $r\times r$ random matrix such that
\begin{align}
\label{Gam}
\gamma_{i,j}=\E\left(\psi'(\epsilon_1)\right)\int_0^1\mathcal{S}_{r-(i-1)}(t)\mathcal{S}_{r-(j-1)}(t)dt \ \ \mbox{for} \ i,j=1,2,\ldots,r.
\end{align}
Moreover, note that
$$
J_n\sum_{t=r+1}^{n} \mathbf{u_{t-1}}\psi(\epsilon_t) \ \overset{d}\to\  \mathcal{F},$$
where
\begin{align}
\label{LJN}
\scriptsize{
 \mathcal{F}= {\E}^{1/2}\left(\psi^2(\epsilon_1)\right)\left[
   \int_0^1\mathcal{S}_{r}(t)dW(t),
  \int_0^1\mathcal{S}_{r-1}(t)dW(t),
   \ldots,
   \int_0^1\mathcal{S}_{1}(t)dW(t)
\right]^\intercal}.
\end{align}
Now, summarize the results to obtain the following limiting distribution
\begin{align}
\label{Lim1}
\small
\left(
J_n^{\intercal}
\right)^{-1}
   \left(\sum_{t=r+1}^{n} \mathbf{u_{t-1}}\mathbf{u_{t-1}}^\intercal\psi'(\epsilon_t)\right)^{-1}\sum_{t=r+1}^{n} \mathbf{u_{t-1}}\psi(\epsilon_t)
    &\ \overset{d}\to \
  \Gamma^{-1}\mathcal{F},
\end{align}
where $\Gamma=(\gamma_{i,j})_{r\times r}$ and $\mathcal{F}$, respectively, are defined in \eqref{Gam} and \eqref{LJN}.
$\hfill\square$
\\

\noindent {\textbf{Proof of Theorem 1 with Roots equal to -1.}} Consider the following model
\begin{align*}
(1+B)^sv_{t}=\epsilon_{t}.
\end{align*}
The limiting distribution in this case is similar to the case in which the time series has unit root 1, except that $\epsilon_{i}$ is replaced by $(-1)^i\epsilon_{i}$ for $i=1,2,\ldots,t$. Similar to Chan and Wei (1988), we define
\begin{align}
\label{newdif-1}
v_{t}(j)=(1+B)^{s-j}v_{t} \ \ \mbox{for} \ j=1,2,\ldots,s.
\end{align}
 Notice that \eqref{newdif-1} implies that $(-1)^{t}v_{t}(j+1)=\sum_{i=1}^{t}(-1)^{i}v_{t}(j)$. By \eqref{3} and \eqref{5}, we have
$$a_n^{-1}(-1)^{t}v_{[nt]}(1)=S_{n}^{(1)}(t)=:\mathcal{S}_{1,n}^{(1)}(t)\ \overset{d}\to \ S^{(1)}(t)=:\mathcal{S}_{1}^{(1)}(t),$$
which implies that
$$a_{n}^{-1}n^{-(j-1)}(-1)^{t}v_{t}(j)=\int_{0}^{t}\mathcal{S}_{j-1,n}^{(1)}(s)ds=:\mathcal{S}_{j,n}^{(1)}(t)\ \overset{d}\to \ \mathcal{S}_{j}^{(1)}(t)$$
for $j=2,\ldots,s$.  Then, let
\begin{align}
\label{KN}
K_n=\aleph_n^{-1}\c{C},
\end{align}
where
\begin{align*}
\scriptsize{
\c{C}=
\left(
 \begin{array}{c c c c c}
    1 & 0 & 0 & \cdots & 0\\
    1 & 1& 0 & \cdots & 0\\
   \vdots & \vdots& \vdots & \ddots & \vdots\\
    1 & \binom{s-1}{1} & \binom{s-1}{2} & \cdots & 1\\
     \end{array}
\right)}
\end{align*}
and the appropriate normalizing constant is $\aleph_n=\text{diag}\left(n^{s-1/2}a_{n}, n^{(s-1)-1/2}a_{n},\ldots,n^{1/2}a_{n}\right)$.
Therefore, similar to the case with root 1, we have
\begin{align*}
 K_n\left(\sum_{t=s+1}^{n} \mathbf{v_{t-1}}\mathbf{v_{t-1}}^\intercal\psi'(\epsilon_t)\right)K_n^{\intercal} &\ \overset{d}\to \ \Upsilon,
\end{align*}
where $\Upsilon=(\upsilon_{i,j})$ is a $s \times s$ random matrix whose entries are as follows:
\begin{align}
\label{Ups}
\upsilon_{i,j}=\E\left(\psi'(\epsilon_1)\right)\int_0^1\mathcal{S}_{s-(i-1)}^{(1)}(t)\mathcal{S}_{s-(j-1)}^{(1)}(t)dt \ \ \mbox{for} \  i,j=1,2,\ldots,s.
\end{align}
We also have
\begin{align*}
\small{
K_n\sum_{t=s+1}^{n} \mathbf{v_{t-1}}\psi(\epsilon_t)\ \overset{d}\to \ \mathcal{H},}
\end{align*}
where
\begin{align}
\label{H1}
 \mathcal{H}= -{\E}^{1/2}\left(\psi^2(\epsilon_1)\right)\left[
   \int_0^1\mathcal{S}_{s}^{(1)}(t)dW(t),
  \int_0^1\mathcal{S}_{s-1}^{(1)}(t)dW(t),
   \ldots,
   \int_0^1\mathcal{S}_{1}^{(1)}(t)dW(t)
\right]^\intercal.
\end{align}
Finally, we get
\small{
\begin{align}
\label{Lim2}
\left(
K_n^{\intercal}
\right)^{-1}
\left(\sum_{t=s+1}^{n} \mathbf{v_{t-1}}\mathbf{v_{t-1}}^\intercal\psi'(\epsilon_t)\right)^{-1}\sum_{t=s+1}^{n} \mathbf{v_{t-1}}\psi(\epsilon_t)\ \overset{d}\to \
  \Upsilon^{-1}\mathcal{H},
\end{align}}
where $\Upsilon=(\upsilon_{i,j})_{s\times s}$ and $\mathcal{H}$ are defined in \eqref{Ups} and \eqref{H1}, respectively.
$\hfill\square$
\\

\noindent {\textbf{Proof of Theorem 1 with Complex Conjugate Unit Roots.}} Consider the model
\begin{align*}
(1-2\cos\theta B+B^2)^dw_{t}=\epsilon_{t}.
\end{align*}
Similar to Chan and Wei (1988), let $y_t(j)=(1-2\cos\theta B+B^2)^{d-j}w_t$ for $j=1,2,\ldots,d$ and $Y_t=(y_t(1),y_{t-1}(1),\ldots,y_{t-1}(d),y_{t-1}(d))^\intercal$. Therefore, we have $(1-2\cos\theta B+B^2)y_{t}(j+1)=y_{t}(j)$, and
\begin{align*}
y_{t}(j+1)= \frac{1}{\sin\theta}\sum_{k=1}^{t}\sin \left(\theta(t-k+1)\right)y_{k}(j) \ \ \mbox{for} \ \ j=0,1,\ldots,d-1.
\end{align*}
We can find a $2d\times 2d$ matrix $D$ such that $D\mathbf{w}_t=Y_t$, where $\mathbf{w}_t=(w_t,\ldots,w_{t-2d+1})$. For more details see Chan and Wei (1988). By applying trigonometric identities, we have
\begin{align}
\label{expx}
\sin(\theta) y_{t}(j)
=a_n\sin\left((t+1)\theta\right) T_{1,t}(j-1)-a_n\cos\left((t+1)\theta\right) T_{2,t}(j-1),
\end{align}
where $T_{1,t}(j)=a_n^{-1}\sum_{k=1}^{t}\cos(k\theta)y_k(j)$ and $T_{2,t}(j)=a_n^{-1}\sum_{k=1}^{t}\sin(k\theta)y_k(j)$. Note that by Lemma  \ref{Asy.Tn2}, we have
\begin{align}
\label{TL}
\left({T_{1,t}(0)},{T_{2,t}(0)}\right)^{\intercal}\overset{d}{ \to}\mathbf{T}(\cdot)=(T_1(t),T_2(t))^\intercal,
\end{align}
where $\mathbf{T}(\cdot)$ is defined in Lemma  \ref{Asy.Tn2}. Now, consider the following representation:
{\scriptsize
\begin{align*}
D&(\sum_{t=2d+1}^{n} \mathbf{w_{t-1}}\mathbf{w_{t-1}}^\intercal\psi'(\epsilon_t))D^\intercal=\left(
 \begin{array}{l l l}
    \sum_{t=2d+1}^{n} y_{t}^2(1)\psi'(\epsilon_t) & \cdots &\sum_{t=2d+1}^{n} y_{t}(1)y_{t-1}(d)\psi'(\epsilon_t)\\
        \vdots &  \ddots & \vdots\\
    \sum_{t=2d+1}^{n} y_{t-1}(d)y_{t}(1)\psi'(\epsilon_t)& \cdots &\sum_{t=2d+1}^{n} y_{t-1}^2(d)\psi'(\epsilon_t)
 \end{array}
\right).
\end{align*}}
To find the limiting distribution of $D\left(\sum_{t=2d+1}^{n} \mathbf{w_{t-1}}\mathbf{w_{t-1}}^\intercal\psi'(\epsilon_t)\right)D^\intercal$,
by using Lemma 3.3.6. of Chan and Wei (1988) and from Proposition 8 of Jeganathan (1991), we have
\begin{align}
\label{ijeg}
\sup\limits_{0<j\leq n} \left|\sum_{k=1}^{j}e^{ik\theta}T_{i,n}\left(\frac{k}{n}\right)\right|=o_p(n)
\end{align}
for $i=1,\ldots, d$. Moreover, by letting
\begin{align}
\label{LN}
L_n=M_n^{-1}D,
\end{align}
where $M_n=\text{diag}\left(n^{1/2}a_{n}I,\ldots,n^{(d-1)/2}a_{n}I\right)$ and  $I=\text{diag}\left(1,1\right)$, we have
\begin{align*}
 L_n\left(\sum_{t=2d+1}^{n} \mathbf{w_{t-1}}\mathbf{w_{t-1}}^\intercal\psi'(\epsilon_t)\right)L_n^{\intercal} &\ \overset{d}\to \ \Lambda.
\end{align*}
Here $\Lambda=(\lambda_{i,j})$ is a $2d\times 2d$ random matrix, where $\lambda_{2i-1,2j-1}=\lambda_{2i,2j}$  and $\lambda_{2i-1,2j}=\lambda_{2i,2j-1}$ for $i,j=1,2,\ldots,2d$. By using \eqref{TL} and \eqref{ijeg} along with the continuous mapping theorem, we can show that $\lambda_{i,j}$ is presented as follows:
{\small
\begin{align}
\label{LCA}
\nonumber\lambda_{2i-1,2j-1}=\lambda_{2i,2j}&=\frac{{\E}\left(\psi'(\epsilon_1)\right)}{2}\times\left(\int_{0}^{1}T_{1,t}(i-1)T_{1,t}(j-1)d(t)+ \int_{0}^{1}T_{2,t}(i-1)T_{2,t}(j-1)d(t)\right),\\
\nonumber\lambda_{2i-1,2j}=\lambda_{2i,2j-1}&=\frac{ {\E}\left(\psi'(\epsilon_1)\right)}{2}\\
\nonumber&\times\left[\cos\theta\left(\int_{0}^{1}T_{1,t}(i-1)T_{1,t}(j-1)d(t)+\int_{0}^{1}T_{2,t}(i-1)T_{2,t}(j-1)d(t)\right)\right.\\
&\ \ \left.-\sin\theta\left(\int_{0}^{1}T_{1,t}(i-1)T_{2,t}(j-1)d(t)- \int_{0}^{1}T_{1,t}(j-1)T_{2,t}(i-1)d(t)\right)\right].
\end{align}}
Also, we have
\begin{align*}
L_n\sum_{t=2d+1}^{n} \mathbf{w_{t-1}}\psi(\epsilon_t) \ \overset{d}\to \mathcal{G},
\end{align*}
where $\mathcal{G}=(\mathcal{G}_{1},\mathcal{G}_{2},\ldots,\mathcal{G}_{2d})^\intercal$. Note that $\mathcal{G}_i$ for $i=1,\ldots,2d$  can be expressed as follows:
{\small
\begin{align}
\label{GL}
\nonumber\mathcal{G}_{2i-1}&={\E}^{1/2}\left(\psi^2(\epsilon_1)\right)
 \left[\cos\theta\left(\int_{0}^{1}T_{1,t}(j-1)dR_1(t)- \int_{0}^{1}T_{2,t}(j-1)dR_2(t)\right) \right.\\
 \nonumber&\quad\quad\quad\quad\quad\quad\quad\left. +\sin\theta\left(\int_{0}^{1}T_{1,t}(j-1)dR_2(t)+ \int_{0}^{1}T_{2,t}(j-1)dR_1(t)\right)\right],\\
\mathcal{G}_{2i}&={\E}^{1/2}\left(\psi^2(\epsilon_1)\right)\times \left(\int_{0}^{1}T_{1,t}(j-1)dR_1(t)- \int_{0}^{1}T_{2,t}(j-1)dR_2(t)\right) ,
\end{align}}
where $\mathbf{R}=(R_1(\cdot),R_2(\cdot))^\intercal$ is defined in \eqref{DR}. Thus, we have the following result:
\begin{align}
\label{Lim3}
\left(L_n^{\intercal}\right)^{-1}\left(\sum_{t=2d+1}^{n} \mathbf{w_{t-1}}\mathbf{w_{t-1}}^\intercal\psi'(\epsilon_t)\right)^{-1}\sum_{t=2d+1}^{n}\mathbf{w_{t-1}}\psi(\epsilon_t)\ \overset{d}\to \ \Lambda^{-1} \mathcal{G},
\end{align}
where $\Lambda=(\lambda_{i,j})$ and $\mathcal{G}=(\mathcal{G}_{1},\mathcal{G}_{2},\ldots,\mathcal{G}_{2d})^\intercal$ are defined in \eqref{LCA} and \eqref{GL}, respectively.
$\hfill\square$
\\

\noindent {\textbf{Proof of Theorem 1 with cross product terms.}}
Chan and Zhang (2012) show that the limiting distributions of the cases  involving the cross product terms converge to zero in probability. We skip the proof of this case, as it is similar to their proof.

$\hfill\square$

\end{document}